\documentclass[aps,prd,nobibnotes,twocolumn,superscriptaddress,bibliography]{revtex4-2}
\usepackage{amsfonts}
\usepackage{mathrsfs}
\usepackage{amsmath}
\usepackage{color}
\usepackage{graphicx}
\usepackage{bm}
\usepackage{amssymb}
\usepackage{xspace}
\usepackage{epstopdf}
\usepackage{dcolumn}
\usepackage{longtable}
\usepackage{multirow}
\usepackage{float}
\usepackage{comment}
\usepackage{braket}
\usepackage{soul}
\usepackage{multirow}
\usepackage{bbding}
\usepackage{blindtext}
\usepackage{longtable}
\usepackage{xcolor}

\usepackage[colorlinks=true, letterpaper=true, pdfstartview=FitV, linkcolor=blue, citecolor=blue, urlcolor=blue]{hyperref}

\makeatother
\begin{document}
	\title{Encyclopedia of emergent particles in 528 magnetic layer groups and 394 magnetic rod groups}

\author{Zeying Zhang}
\affiliation{College of Mathematics and Physics, Beijing University of Chemical
	Technology, Beijing 100029, China}
\affiliation{Research Laboratory for Quantum Materials, Singapore University of
	Technology and Design, Singapore 487372, Singapore}

\author{Weikang Wu}
\email{weikang\_wu@sdu.edu.cn}
\affiliation{Key Laboratory for Liquid-Solid Structural Evolution and Processing of Materials, Ministry of Education, Shandong University, Jinan 250061, China}

\author{Gui-Bin Liu}
\affiliation{Centre for Quantum Physics, Key Laboratory of Advanced Optoelectronic
	Quantum Architecture and Measurement (MOE), Beijing
	Institute of Technology, Beijing 100081, China}
\affiliation{Beijing Key Lab of Nanophotonics \& Ultrafine Optoelectronic Systems,
	School of Physics, Beijing Institute of Technology, Beijing 100081,
	China}

\author{Zhi-Ming Yu}
\affiliation{Centre for Quantum Physics, Key Laboratory of Advanced Optoelectronic
	Quantum Architecture and Measurement (MOE), Beijing
	Institute of Technology, Beijing 100081, China}
\affiliation{Beijing Key Lab of Nanophotonics \& Ultrafine Optoelectronic Systems,
	School of Physics, Beijing Institute of Technology, Beijing 100081,
	China}

\author{Shengyuan A. Yang }
\email{shengyuan\_yang@sutd.edu.sg}
\affiliation{Research Laboratory for Quantum Materials, Singapore University of
	Technology and Design, Singapore 487372, Singapore}

\author{Yugui Yao}
\email{ygyao@bit.edu.cn}
\affiliation{Centre for Quantum Physics, Key Laboratory of Advanced Optoelectronic
	Quantum Architecture and Measurement (MOE), Beijing
	Institute of Technology, Beijing 100081, China}
\affiliation{Beijing Key Lab of Nanophotonics \& Ultrafine Optoelectronic Systems,
	School of Physics, Beijing Institute of Technology, Beijing 100081,
	China}

\begin{abstract}
We present a systematic classification of emergent particles in all
528 magnetic layer groups and 394 magnetic rod groups, which describe two-dimensional and one-dimensional crystals respectively. Our approach is via constructing a correspondence between a given magnetic layer/rod group and one of the magnetic space group, such that all irreducible representations of the layer/rod group can be derived from those of the corresponding space group.
Based on these group representations, we explicitly construct the effective models for possible band degeneracies and identify all emergent
particles, including both spinless and spinful cases. We find that there are six kinds of particles protected by magnetic layer groups and three kinds by magnetic rod groups.
Our work provides a useful reference for the search and design of emergent particles in lower dimensional crystals.

\end{abstract}
\maketitle
	
\section{Introduction}
The research on topological semimetals in the past decade has driven extensive efforts in understanding various emergent particles enabled by the crystalline symmetry \cite{chiu_classification_2016,Burkov2016,armitage_weyl_2018}.
In these crystals, novel kinds of quasi-particle states emerge around band degeneracies in the momentum space, and their physical properties are determined by the character of the band degeneracies. For example, Weyl and Dirac particles can be realized around twofold and fourfold band nodal points in crystals, known as Weyl and Dirac points, respectively \cite{wan_topological_2011,young_dirac_2012,wang_dirac_2012,armitage_weyl_2018}. These notions are not limited to electronic systems of real materials, but also extended to many artificial crystals such as acoustic/photonic crystals \cite{lu_topological_2014,yang_topological_2015,ozawa_topological_2019}, electric circuit arrays \cite{ningyuan_time-_2015,imhof_topolectrical-circuit_2018,yu_4d_2020}, and mechanical networks \cite{ huber_topological_2016,ma_topological_2019}.

A main task in this research is to classify all possible types of emergent particles.
The classification is typically based on the dimension of the degeneracy manifold, the degree of degeneracy, the band dispersion, and the topological charge. For instance, in three dimensions (3D), besides nodal points, the band degeneracies may also form nodal lines \cite{yang_dirac_2014,weng_topological_2015,mullen_line_2015,chen_nanostructured_2015} or nodal surfaces \cite{zhong_towards_2016,liang_node-surface_2016,wu_nodal_2018}; the degeneracy for a nodal point could be 2, 3, 4, 6, and 8 \cite{zhu_triple_2016,bradlyn_beyond_2016,yu_encyclopedia_2022}; the emergent particle may have linear, quadratic or cubic dispersion \cite{fang_multi-weyl_2012,xu_chern_2011, yu_quadratic_2019, zhang_magnetic_2021}; and they could have a maximal chiral charge of four \cite{yu_encyclopedia_2022,cui_charge-four_2021}.
All these properties are eventually determined by the symmetry of the bands that form the degeneracy, or more specifically, how these bands represent the symmetry group of the system. For 3D crystals, the pertinent symmetry groups are the (magnetic) space groups. Their (co-)representations have been extensively studied and well documented in the past \cite{bradley_mathematical_2009} (in the remainder of this letter the ``representation" means representation for unitary group and co-representation for magnetic group). Based on the knowledge of space group representations, an encyclopedia of emergent particles in 3D crystals has been established in recent works \cite{tang_exhaustive_2021,yu_encyclopedia_2022,liu_systematic_2022,zhang_encyclopedia_2022,tang_complete_2022}.

\begin{figure*}[t]
	\includegraphics[width=\linewidth]{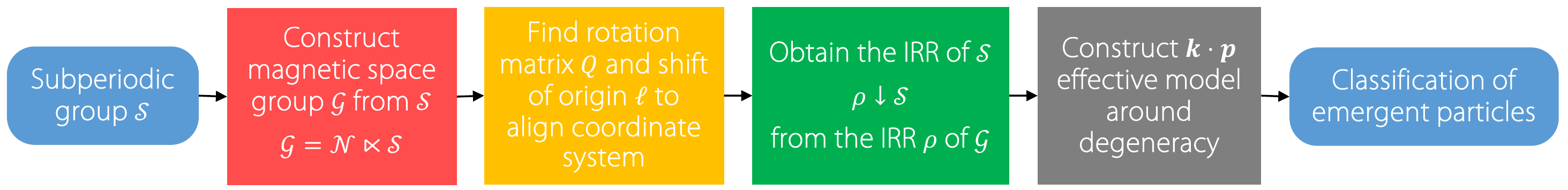}
	\caption{Flow chart for classification of emergent particles in subperiodic groups.}
	\label{fig:flow}
\end{figure*}


Recent years also witnessed a rapid development in the realization of low-dimensional crystals. Many 2D layered materials and 1D (or quasi-1D) crystals have been synthesized in experiment \cite{xia_one-dimensional_2003, miro_atlas_2014,tan_recent_2017}. Because lower dimensions permit a high controllability and easier detection,
emergent particles may have even stronger impact in these systems. As a prominent example, many intriguing properties of graphene can be attributed to its emergent Dirac fermions \cite{castro_neto_electronic_2009}. So far, there have been works on studying specific kind of emergent particles in 2D~\cite{feng_two-dimensional_2021} and on systematic construction of $k\cdot p$ models for 2D systems \cite{park_classification_2017, tang_complete_2022, lazic_fully_2022}. However, a comprehensive classification for all emergent particles in magnetic layer groups (MLGs) and magnetic rod groups (MRGs), which apply for 2D and 1D crystals respectively, has not been achieved.

One obstacle here is that the irreducible representations (IRRs) have not been completely derived for these groups, but only for certain subgroups (such as the type-I and type-II MLGs) \cite{litvin_character_1991,nikolic_irreducible_2022}. In this work, we develop an approach to compute IRRs for all 528 MLGs and 394 MRGs. The approach is based on making a correspondence between a given MLG/MRG and one of the magnetic space groups. We show that all IRRs of the MLG/MRG can be derived by restricting the IRRs of the corresponding space group under a constraint. After obtaining IRRs for these groups, we identify all possible protected band degeneracies and classify the associated emergent particles, for both spinless and spinful systems.  We find six kinds of emergent particles in MLGs and three kinds in MRGs, as listed in Table~I. Our work offers a comprehensive reference for the investigation of emergent particles in low dimensional crystals.

\section{Derive IRRs of subperiodic groups}
{MLGs and MRGs are subperiodic groups in 3D, meaning that their translational parts form a 2D or 1D subspace of 3D. Each of their point groups remains one of the 3D crystallographic point groups. To derive their IRRs, one can of course pursue the standard way as in Ref.~\cite{bradley_mathematical_2009}, e.g., by studying little co-groups and using the method of projective  representations.  Here, we shall adopt an alternative approach, in which we obtain the IRRs of a magnetic subperiodic group from a constructed magnetic space group.


\subsection{General approach}
Our approach is a unified treatment for both MLGs and MRGs. Consider a given subperiodic group $\mathcal{S}$ \cite{litvin_magnetic_2016}. We first construct a magnetic space group $\mathcal{G}$ from $\mathcal{S}$. This is done by taking a lattice translation group $\mathcal{N}$, which consists of translations normal to the subspace for $\mathcal{S}$. For a MLG, $\mathcal{N}$ is the 1D lattice translations normal to the plane. For a MRG, $\mathcal{N}$ contains the 2D translations normal to the line. Then the space group $\mathcal{G}$ is constructed as the semidirect product:
\begin{equation}
  \mathcal{G}=\mathcal{N}\rtimes \mathcal{S}.
\end{equation}
In the Supplemental Material (SM), we present the chosen $\mathcal{G}$ for each MLG and MRG in Table~S1 \cite{noauthor_notitle_nodate}.

By definition, $\mathcal{N}$ is a normal subgroup of $\mathcal{G}$, and $\mathcal{S}$ is isomorphic to $\mathcal{G}/\mathcal{N}$, with the isomorphism
\begin{equation}
  \phi:   s\in \mathcal{S} \mapsto \mathcal{N}s\in \mathcal{G}/\mathcal{N}.
\end{equation}
Thus, to get IRRs of $\mathcal{S}$ is equivalent to obtain the IRRs of the quotient group $\mathcal{G}/\mathcal{N}$.

Since $\mathcal{G}$ is one of the magnetic space groups, its IRRs are already known (here obtained by using the \textsf{MSGCorep} package) \cite{liu_spacegroupirep_2021,liu_msgcorep_2021}. With this information, all IRRs of
$\mathcal{S}$ can be obtained from restricting IRRs of $\mathcal{G}$ to $\mathcal{S}$. However, not all IRRs of  $\mathcal{G}$
lead to IRRs of $\mathcal{S}$ under restriction. For some IRRs of  $\mathcal{G}$, the restriction to its subgroup would lead to reducible representations. Then, what are the suitable IRRs $\rho$ of $\mathcal{G}$ that we need to consider? From group representation theory, these are the IRRs which satisfy the condition that $\ker \rho$ contains $\mathcal{N}$ as a subgroup~\cite{james_representations_2001}, i.e.,
\begin{equation}\label{N}
  \mathcal{N}\subseteq \ker \rho=\{g\in \mathcal{G}|\rho(g)\text{ is identity matrix}\}.
\end{equation}
In other words, the normal translations must be represented as identity matrix in the IRR $\rho$ of $\mathcal{G}$. The restriction of such $\rho$ to $\mathcal{S}$, i.e., $\rho\downarrow \mathcal{S}$, is indeed an IRR of $\mathcal{S}$. This can be easily verified from the corresponding restricted character $\chi\downarrow \mathcal{S}$, where $\chi$ is the character of $\rho$.
Recall that a representation $\rho$ of $G$ is an IRR if and only if its character satisfies $\braket{\chi,\chi}_{G}=1$,  $\braket{\theta,\phi}_G$ denotes the inner product of characters for group $G$. Now, the restricted character $\chi\downarrow \mathcal{S}$ satisfies
\begin{equation}
	\begin{split}
		\braket{\chi\downarrow \mathcal{S},\chi\downarrow \mathcal{S}}_{\mathcal{S}}
		=&\frac{1}{|\mathcal{S}|}\sum_{s\in \mathcal{S}}\chi(s)\chi^*(s)\\
		=&\frac{1}{|\mathcal{S}||\mathcal{N}|}\sum_{n\in \mathcal{N}} \sum_{s\in \mathcal{S}}\chi(ns)\chi^*(ns)\\
		=&\frac{1}{|\mathcal{G}|}\sum_{g\in \mathcal{G}}\chi^*(g)\chi(g)=\braket{\chi,\chi}_{\mathcal{G}},\\
	\end{split}
\end{equation}
where in the second step, we used the
fact that $\chi(ns)=\chi(s)$ since $\mathcal{N}\subseteq \ker \rho$. 
Thus, the restricted representation $\rho\downarrow \mathcal{S}$ for $\mathcal{S}$ shares the same irreducibility as $\rho$ for $\mathcal{G}$.

In a similar way, one can show that the indicator function for such $\rho$ is also preserved in the process of restriction to $\mathcal{S}$, i.e.,
$\sum_{s\in \mathcal{S}_A}(\chi\downarrow \mathcal{S})(s^2)=\sum_{g\in \mathcal{G}_A}\chi(g^2)$, where $\mathcal{S}_A$ and $\mathcal{G}_A$ are anti-unitary part of $\mathcal{S}$ and $\mathcal{G}$  respectively. Thus, the restricted IRR for $\mathcal{S}$ shares the same type of corepresentation as the original IRR for $\mathcal{G}$.

\subsection{Algorithm for aligning coordinate systems}

\begin{table*}[t]
	\caption{Emergent particles in MLGs. ``$\surd$'' ( ``$\times$'') means the corresponding emergent particle can (cannot) exist in the specified type of magnetic subperiodic groups. $d_m$ is the dimension of the degeneracy manifold. $d$ is the degree of degeneracy. Ld is the leading order dispersion of the band splitting near the degeneracy. ``I'' to ``IV'' represent the four types of magnetic subperiodic groups.
	}
	\label{tab:mlgsum}
	\begin{ruledtabular}
		\begin{tabular}{llllcllll}
			Notation&Abbr.&$d_m$&$d$&Ld&I & II & III & IV \\
			\hline	
			\textbf{MLGs w/o spin}  	&&&&& &  &  &  \\
			Linear Weyl point&LWP				&0&2&(11)&$\surd$&$\surd$&$\surd$&$\surd$\\
			Quadratic Weyl point& QWP		&0&2&(22)&$\surd$&$\surd$&$\surd$&$\surd$\\
			Dirac point&	DP			&0&4&(11)&$\times$ &  $\surd$ & $\surd$ & $\surd$ \\
			Weyl line&WL				&1&2&(1)&$\surd$&$\surd$&$\surd$&$\surd$\\
			Weyl line net&WLs			&1&2&(1)&$\surd$&$\surd$&$\surd$&$\surd$\\		
			&&& && &  &  & \\
			\textbf{MLGs w/ spin}		&&& && &  &  &    \\
			Linear Weyl point&LWP				&0&2&(11)&			$\surd$&$\surd$&$\surd$&$\surd$\\
			Quadratic Weyl point& QWP		&0&2&(22)&	$\surd$&$\times$&$\surd$&$\times$\\
			Cubic Weyl point&	CWP 			&0&2&(33)&	$\surd$&$\surd$&$\times$&$\times$\\
			Dirac point&	DP 			&0&4&(11)&	$\surd$&$\surd$&$\surd$&$\surd$\\
			Weyl line&WL				&1&2&(1)&	$\surd$&$\surd$&$\surd$&$\surd$\\
			Weyl line net&WLs			&1&2&(1)&	$\surd$&$\surd$&$\surd$&$\surd$\\
			Dirac line&DL		&1&4&(1)&	$\times$&$\surd$&$\surd$&$\surd$\\
			Dirac line net&DLs	&1&4&(1)&	$\times$&$\surd$&$\surd$&$\surd$\\
		\end{tabular}
	\end{ruledtabular}
\end{table*}

It is not difficult to identify the IRRs of $\mathcal{G}$ that satisfy condition (\ref{N}). For example, for $\mathcal{S}$ describing a 2D system in the $x$-$y$ plane, one can easily see that the required IRR for $\mathcal{G}$ should correspond to $k_z=0$ plane of the 3D Brillouin zone (BZ). Similarly, for a MRG describing a 1D system along $z$, the IRR for $\mathcal{G}$ should correspond to $k_x=k_y=0$ path of the BZ.

There is another technical issue arising in practical calculations. Usually, the coordinate system for the standard setting of a magnetic space group, as in well-known references, is different from the $\mathcal{G}$ constructed here. The two generally differs by a proper rotation and a shift of origin. Thus, in order to use the documented IRRs for magnetic space groups, we need to align the two coordinate systems.

Let $\mathsf{L}=(\boldsymbol{a},\boldsymbol{b},\boldsymbol{c})$ be lattice vectors for the $\mathcal{G}$ constructed from $\mathcal{S}$ and $\mathsf{L}'=(\boldsymbol{a}',\boldsymbol{b}',\boldsymbol{c}')$ be lattice vectors in the standard setting for this space group.
The two set of lattice vectors differ by a linear transformation $Q$:
\begin{equation}\label{LL}
	\mathsf{L}=\mathsf{L}'Q,
\end{equation}
where $Q$ is a $3\times3$ matrix and $\det Q =1$ and is determined by $Q=\mathsf{L}\mathsf{L}'^{-1}$.

Besides the rotation, the two coordinate systems may have shift in origin. Consider a general point labeled by coordinate $\bm x$ in the $\mathsf{L}$ system. Its coordinate in $\mathsf{L}'$ is given by
\begin{equation}
	\boldsymbol{x}'=Q\boldsymbol{x}+\bm\ell,
\end{equation}
where $\bm\ell$ is the shift of origin between the two coordinate system.
This shift enters into the expression of general space group symmetry operation.  Considering the expression of any space group symmetry operation, we should have
\begin{equation}\label{RR}
	\begin{split}	
    \{R'|\boldsymbol{\tau}'\}&=\big\{Q RQ^{-1}|Q\boldsymbol{\tau}-QRQ^{-1}\bm{\ell}+\bm\ell\big\}.
	\end{split}
\end{equation}

In our current case, we already know the lattice vectors $\mathsf{L}$ and $\mathsf{L}'$, and the expressions $\{R_i|\bm \tau_i\}$ and $\{R_i '|\bm \tau_i '\}$ ($i$ labels the symmetry generators). The target is to solve out $\bm \ell$.
This is done by using the following algorithm.

We write down the following set of equations obtained from (\ref{RR}):
\begin{equation}\label{8}
	\begin{split}
		\bm \tau_i'&=Q\bm\tau_i-QR_{i}Q^{-1}\bm\ell+\bm\ell \pmod{1},
	\end{split}
\end{equation}
where the lattice periodicity allows the two sides differing by a lattice period. Note that
this set of equations are not conventional linear congruence equations, since $\bm\tau_i', \bm\tau_i$ and $\bm\ell$ can be fractional numbers. We can multiply Eq.~(\ref{8}) by integer $N$ to turn them into linear congruence equations
\begin{equation}
	\begin{split}\label{9}
		N\bm \tau_i'&=Q(N\bm\tau_i)-QR_{i}Q^{-1}(N\bm\ell)+(N\bm\ell) \pmod{N},
	\end{split}
\end{equation}
where $N$ is the least common multiple of the denominators of $\bm\tau_i', \bm\tau_i$ and $\bm\ell$.
Here, although we do not yet know the denominator of $\bm\ell$, in space groups, it will not be a large integer. In fact, from our calculation, the maximum $N$ is just 12. Therefore, in practice, one may just try to increasingly select a $N$, solve $\bm\ell$ from (\ref{9}) by using the Chinese remainder theorem \cite{artin_algebra_2011}, and check whether such a solution is valid.

The whole process for obtaining the IRRs of a subperiodic group $\mathcal{S}$ is schematically illustrated in Fig.~\ref{fig:flow}.

\begin{table*}[t]
	\caption{Emergent particles in MRGs. The format of this table is similar to Table~\ref{tab:mlgsum}.
	}
	\label{tab:mrgsum}
	\begin{ruledtabular}
		\begin{tabular}{llllcllll}
			Notation&Abbr.&$d_m$&$d$&Ld&I & II & III & IV \\
			\hline
			\textbf{MRGs w/o spin} 		&&&&& &  &  &  \\
			Weyl point&WP				&0&2&(1)&$\surd$&$\surd$&$\surd$&$\surd$\\
			Triple point&	TP			&0&3&(1)&	$\surd$&$\surd$&$\surd$&$\surd$\\
			Dirac point&	DP 			&0&4&(1)& $\surd$ &  $\surd$ & $\surd$ & $\surd$ \\
			&&& && &  &  & \\
			\textbf{MRGs w/ spin}		&&& &  &&&  &    \\				
			Weyl point&WP				&0&2&(1)&$\surd$&$\surd$&$\surd$&$\surd$\\
			Triple point&	TP 			&0&3&(1)& 	$\surd$&$\surd$&$\surd$&$\surd$\\
			Dirac point&	DP 			&0&4&(1)&$\surd$ &  $\surd$ & $\surd$ & $\surd$\\
		\end{tabular}
	\end{ruledtabular}
\end{table*}

\section{Classification of emergent particles}

After deriving the IRRs of a given MLG or MRG, we can use them to identity possible band degeneracies and the associated emergent particles. The remaining steps are exactly the same as our previous works in Refs.~\cite{liu_systematic_2022,zhang_encyclopedia_2022} on the classification for magnetic space groups.

For emergent particles around a degeneracy point $\bm k$ in BZ. We construct the $k\cdot p$ effective models constrained by symmetry conditions:
\begin{equation}
	H(\boldsymbol{k})=
	\begin{cases}
		D(S)H(R^{-1}\boldsymbol{k})D^{-1}(S),& \text{if } S = \{R|\bm\tau\}\\
		D(S)H^*(-R^{-1}\boldsymbol{k})D^{-1}(S),& \text{if } S = \{R|\bm\tau\}{\cal T},\\
	\end{cases}
\end{equation}
where the rotation part of $S$ will run through all symmetry generators in the magnetic little co-group at $\bm k$, $D$ is its representation corresponding to the band degeneracy, and $\mathcal{T}$ is time reversal operation. We have developed a general algorithm to construct such effective models and implemented it in the \textsf{MagneticKP} package, as introduced in Ref.~\cite{zhang_magnetickp_2022}

\begin{table}[t]
	\caption{List of type-II MLGs that host each kind of emergent particles.}
	\label{tab:mlgii}
	\begin{ruledtabular}	
	\begin{tabular}{ll}
		Name & Layer groups \\
		\hline
		\textbf{w/o spin} &\\
		LWP&8-10,14-48,53-64,66-73,75-80 \\
		QWP&49-80\\
		DP&29,33,40,43-45,63\\
		WL&5,7,9,12,15-17,20-21,24-25,27-48,51-52,54,56,58\\
		& 60-64,74-75,78-80\\
		WLs &7,21,25,32,34,39,42,44,46,52,54,56,58,60,62-64\\
		&\\
	\textbf{w/ spin} &\\
		LWP& {\color{black} 1,3,5,8-13,19-26,31-34,36,49-50,53-60,65,67-70}\\
& {\color{black} 73,76-77}\\
CWP&{\color{black} 65,67-70,73,76-77}\\
DP&{\color{black} 7,15-17,21,25,28-30,32-34,38-39,41-43,45-46,48,52}\\
& {\color{black}54,56,58,60,62,64} \\
WL&{\color{black} 4-5,9,12,17,20-21,24-25,27-36,54,56,58,60,74,78-79}\\
WLs&{\color{black} 74,78,79}\\
DL&{\color{black}40,43,44,45,63 }\\
DLs& {\color{black}44,63}\\
	\end{tabular}
	\end{ruledtabular}
\end{table}

\begin{table}[t]
	\caption{List of type-II MRGs that host each kind of emergent particles.}
	\label{tab:mlgiisoc}
	\begin{ruledtabular}	
	\begin{tabular}{ll}
		Name & Rod groups  \\
		\hline
\textbf{w/o spin} &\\		
WP&5,7-9,11-26,28-44,46-58,60-75\\
TP&27-29,34-41,45,49-52,59-61,68-75\\
DP&36,40,50,52,60,61,68-70,72-75\\
&\\
\textbf{w/ spin} &\\	
		WP& {\color{black}1,3-5,8-10,13,14,18,19,23-26,30-33,42-44,46-50,53-59}\\
		& {\color{black}62-67,71,72} \\
		TP& {\color{black}49,50,59,71,72} \\
		DP&{\color{black} 7,12,16,17,21,22,28,29,34-36,38-41,45,50-52,60,61}\\
		&{\color{black}68-70,72-75}\\	
	\end{tabular}
	\end{ruledtabular}
\end{table}

The main results from our classification  are summarized in Table~\ref{tab:mlgsum} (for MLGs) and Table.~\ref{tab:mrgsum} (for MRGs). In the first column, we list the kinds of emergent particles that are protected by MLGs or MRGs. In Refs.~\cite{yu_encyclopedia_2022,liu_systematic_2022,zhang_encyclopedia_2022}, we showed that there are 27 kinds of emergent particles protected by magnetic space groups in 3D. Here, for MLGs and MRGs, due to the reduced symmetry, we find that the variety of particles is also much reduced. In MLGs, there are six kinds of particles, corresponding to (linear) Weyl point, quadratic Weyl point, cubic Weyl point, Dirac point, Weyl line, and Dirac line. In the naming, we use Weyl and Dirac to indicate the number of degeneracy to be 2 and 4, respectively, consistent with the convention in Refs.~\cite{yu_encyclopedia_2022,liu_systematic_2022,zhang_encyclopedia_2022}. For MRGs, there are only three kinds of emergent particles, which correspond to Weyl point, triple point, and Dirac point. Some basic characters of these band degeneracies, such as the dimension, the degree of degeneracy, and the leading order band splitting, are also listed in Table~\ref{tab:mlgsum}-\ref{tab:mrgsum}.

The remaining columns of Table~\ref{tab:mlgsum}-\ref{tab:mrgsum} show the possible appearance of a kind of emergent particles in the four types of MLGs or MRGs.

The type-II MLGs and MRGs describe the nonmagnetic 2D and 1D crystals, which cover most existing materials. Taking them as examples, we list the candidate type-II MLGs and MRGs that host each kind of emergent particles in Table~\ref{tab:mlgii} (for MLGs) and \ref{tab:mlgiisoc} (for MRGs).

In Supplemental Material \cite{noauthor_notitle_nodate}, we present the following detailed information. For each kind of emergent particles, we list all MLGs or MRGs that can host it. For each MLG or MRG, we list the band degeneracies it can have, their locations in BZ, the symmetry generators, their representations, and the effective models. The format of such a dictionary follows the convention set in Ref.~\cite{zhang_encyclopedia_2022}

We have a few remarks before proceeding. First, the well-known double degeneracy for all point in BZ due to the spacetime inversion symmetry $\mathcal{PT}$ for spinful systems is not counted in our classification. Second, most degeneracies occur on high-symmetry points or high-symmetry paths of BZ. However,
there are three mechanisms that can protect Weyl point or Weyl lines at generic points of BZ. (i) Spinless MLGs with $\mathcal{PT}$ symmetry can protect Weyl points at generic $k$ points. (ii) Spinless or spinful MLGs with  $C_{2z}\mathcal{T}$ symmetry can protect Weyl points at generic $k$ points. (iii)  MLGs with a horizontal mirror plane can protect Weyl lines passing through generic $k$ points. For these cases, one needs to carefully scan the BZ when looking for band degeneracies.

\section{Discussion and Conclusion}


It is worth mentioning that the spinful quadratic Weyl point, which was found not existing in type-II MLGs, can exist in  type-I and type-III MLGs, as shown in Table~I. For example, consider the type-III MLG $p\bar6'$ (No.~74.3.494). The co-representation matrices for two symmetry generators $C_{3z}$ and $\sigma_h \mathcal{T}$ in the basis of $\Gamma_4\Gamma_5$ degeneracy are
\begin{equation}
  D(C_{3z})=e^{i \pi \sigma_3/3},\qquad  D(\sigma_h \mathcal{T})=\sigma_1,
\end{equation}
where $\sigma_i$'s are the Pauli matrices. Then the effective model for states around the $\Gamma_4\Gamma_5$ degeneracy is
\begin{equation}
	H(\boldsymbol{k})=c_1(k_x^2+k_y^2)+\big[(c_2k_{-}^2+c_3k_{+}^2)\sigma_++\text{H.c.}\big],
\end{equation}
where the momentum and the energy are measured from the degeneracy point, and the $c_i$'s are real-valued parameters. This model confirms that the degeneracy point is a quadratic Weyl point in a spinful system.
Furthermore, for MLG $p\bar6'$, this quadratic Weyl point can be the only Fermi point of a band structure. To show this, we construct a tight-binding model on a hexagonal lattice as shown in Fig.~\ref{fig:qp}(a,b). Here, each active site (marked by red color) has two $s$-like orbitals $\{\ket{s\uparrow}, \ket{s\downarrow}\}$. The grey colored sites are added to enforce the proper MLG, but they do not contain active orbitals. We construct the following model that satisfy the $p\bar6'$ symmetry \cite{zhang_magnetictb_2022}:
\begin{equation}
	\label{eq:qp}
H(\boldsymbol{k})=\varepsilon+\begin{pmatrix}
	\bm \alpha_1 \cdot (t_1 \bm  A +t_2\bm  B )& \bm \alpha_2\cdot t_3\bm A\\
	\dagger & \bm \alpha_1 \cdot (t_1 \bm A -t_2 \bm B) \\
\end{pmatrix},
\end{equation}
where the bold symbols are vectors, $\bm\alpha_1=(1,1,1)$, $\bm \alpha_2=(e^{-\frac{i\pi}{6}},-e^{\frac{i\pi}{6}},i)$, $\bm A=(\cos k_a,\cos k_b,\cos (k_a+k_b))$, $\bm B=(\sin k_a,\sin k_b,\sin (k_a+k_b))$, $k_a$ and $k_b$ are wave vector components along the two reciprocal lattice vectors, and $\varepsilon,t_1,t_2,t_3$ are real parameters. The resulting band structure in Fig.~\ref{fig:qp}(b,c) demonstrates our claim. Interestingly, when one adds an extra horizontal mirror $\sigma_h$ to the system, i.e., the type-III MLG $p\bar6'$ is turned into the type-II MLG $p\bar61'$ The Weyl point at $\Gamma$ will no longer be an isolated nodal point, but sit on the intersection of three nodal loops, as illustrated in Fig.~\ref{fig:qp}(d)

\begin{figure}[t]
	\includegraphics[width=\linewidth]{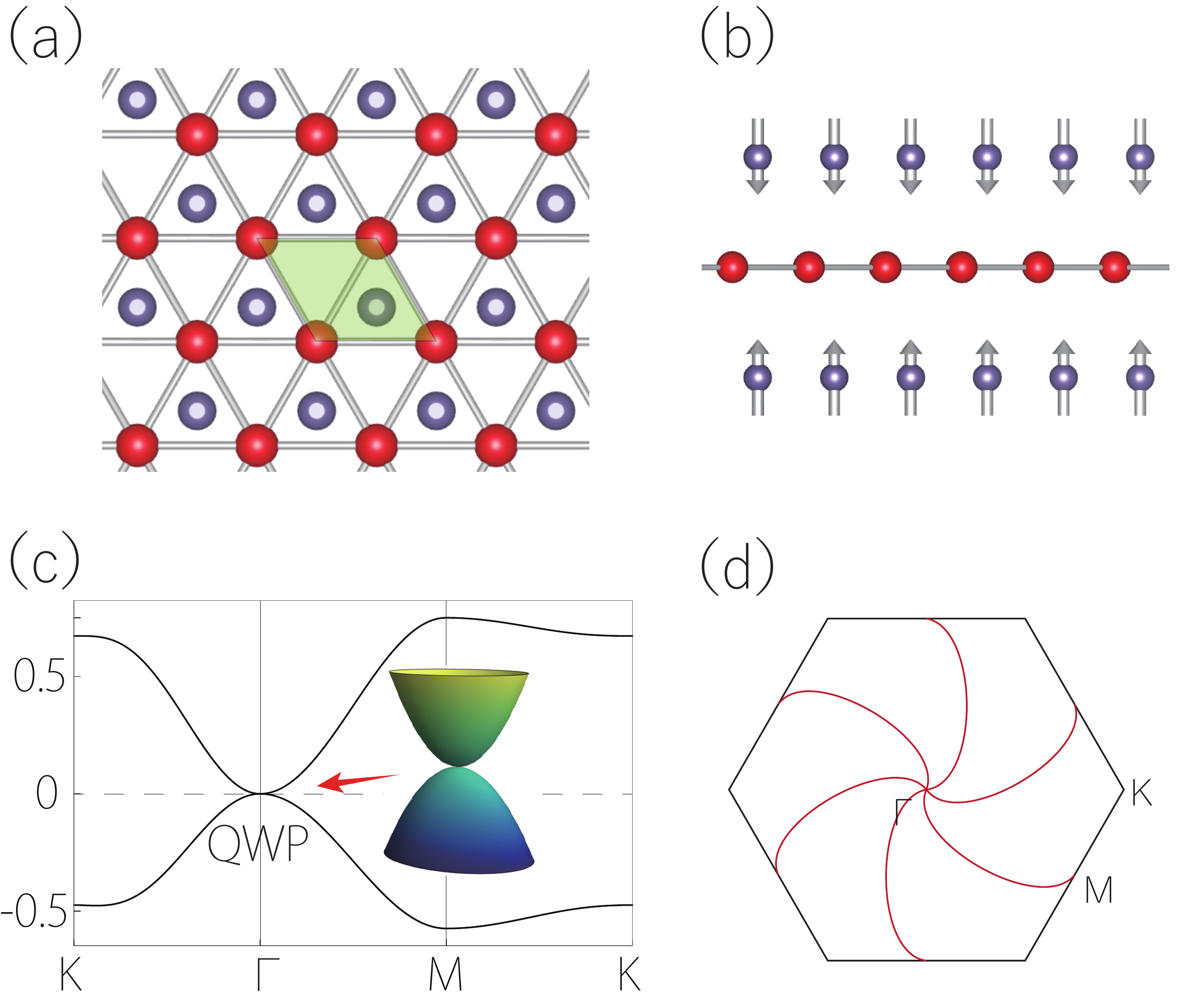}
	\caption{(a) Top view and (b) side view of our constructed QWP lattice model. The green shaded region indicates
		the unit cell. The active orbitals are at the red colored sites. The grey colored sites with local magnetic moments are added to enforce the proper MLG symmetry. (c) Band structure of the model (\ref{eq:qp}). In the calculation, we take $\varepsilon=0.06,\ t_1=-0.02,\ t_2=0.2,\ t_3=-0.3$.  Inset shows the dispersion around the QWP.
		(d) After adding a horizontal mirror to the model, the MLG changes to the type-II MLG $p\bar{6}1'$. Then there appears three WLs as indicated by the red lines.}
	\label{fig:qp}
\end{figure}

{We find that the cubic Weyl point can exist in spinful type-II MLGs, consistent with the prediction in Ref.~\cite{wu_higher-order_2021}. In addition, our result shows that it also exists in spinful type-I MLGs, but not in type-III and type-IV MLGs. This can be understood from the following analysis. A cubic Weyl point necessarily requires one of these six symmetry groups generated by:  $\{C_3,\mathcal{T}\}$,$\{C_6,\mathcal{T}\}$, $\{C_6,C_{21}\}$, $\{C_6,C_{21},\mathcal{T}\}$,
$\{C_6,\sigma_{d}\}$ and $\{C_6,\sigma_{d},\mathcal{T}\}$ respectively,  where $C_{21}$ and $\sigma_{d}$ are rotation and mirror  perpendicular to the $C_6$ axis. It turns out that the index $2$ subgroup of above six magnetic point groups cannot be a type-III magnetic point group, indicating that the type-III MLGs cannot host a cubic point. Moreover, if a type-IV MLG can host the cubic point,
the combination of a half lattice translation and $\mathcal{T}$ must be compatible with the $C_3$ symmetry. However, the relation $C_3\{\mathcal{T}|\frac{1}{2}0\}C_3^{-1}=\{\mathcal{T}|0\frac{1}{2}\}$ (or $C_3\{\mathcal{T}|\frac{1}{2}\frac{1}{2}\}C_3^{-1}=\{\mathcal{T}|\frac{1}{2}0\}$) will give rise to an invalid symmetry element $\{E|\frac{1}{2}\frac{1}{2}\}$ (or $\{E|0\frac{1}{2}\}$). Therefore, cubic Weyl points also cannot exist in type-IV MLGs. }

Although the number of kinds of emergent particles in MRGs is less than that in MLGs, we find that the triple points can exist in MRGs but not in MLGs.
Triple points have been extensively studied in 3D systems. In magnetic space groups, they may appear as essential degeneracies at high-symmetry points. However, this is possible only in cubic system, and we find that such triple points cannot be maintained when restricting to 2D or 1D subperiodic systems.
Another possibility to form a triple point is by crossing a doubly degenerate band with a non-degenerate band, corresponding to a direct sum of a 1D co-representation and a 2D co-representation on a high symmetry line or plane. In MLGs, this has to be on a high-symmetry path of 2D BZ. The little co-group $\mathcal{L}$ on this path should be one from $C_1$, $C_2$, $C_s$, and $C_{2v}$. By using the property of characters
\begin{equation}
	\sum_i\chi_i^2(e)=|\mathcal{L}|,
\end{equation}
where $e$ is the identity element and  the sum is over all IRRs. One finds that for the four little co-groups,
the dimensions of IRRs are  either all 1 or all 2. Therefore, it is impossible to generate a triple point. When taking anti-unitary operations into account, one can show that the dimensions of co-representations are either unchanged or multiplied by 2.
Thus, MLGs cannot host triple points but can host Dirac points and Dirac lines.
By contrast, MRGs can have other rotational symmetries (such as $C_3, C_4, C_6$) in addition to $C_{2}$  along its periodic direction. Therefore, these additional symmetries can enable triple points in MRGs.

It must be noted that MRGs are defined as subperiodic groups of magnetic space groups. They only contain symmetries that are inherited from 3D crystals. However, for 1D or quasi-1D crystals, they may have symmetries that do not exist for 3D crystals. For example, one can easily picture a 1D crystal with $C_5$ symmetry or even $C_\infty$ symmetry. The groups that capture all these possibilities are known as the line groups \cite{damnjanovic_line_2010}. Our study here can be extended to line groups in the near future.

In conclusion, we systematically classify the emergent particles in 528 MLGs and 394 MRGs. In the process, we develop a general approach to derive all IRRs of a subperiodic group by constructing a corresponding superperiodic group and by properly restricting the IRRs of the superperiodic group. This approach is applied to obtain all IRRs of MLGs and MRGs.
Using these IRRs, we established an encyclopedia of emergent particles in MLGs and MRGs. This serves as a valuable reference for the search of novel emergent particles in lower dimensional materials. It can also be used to facilitate the design of artificial crystals to study the fascinating properties of emergent particles.

\begin{acknowledgments}
The authors thank D. L. Deng for helpful discussions. We acknowledge support from  National Key R\&D Program of China (Grant No. 2020YFA0308800), the NSF of China (Grants Nos. 12234003, 12004028, 12004035,  12274028, 52161135108 and 12061131002) and the Singapore MOE AcRF Tier 2 (MOE-T2EP50220-0011). 
W.W. is supported by the Special Funding in the Project of Qilu Young Scholar Program of Shandong University.
\end{acknowledgments}

\textit{Note added.} 
{\color{black} \bf The Supplemental Material of this manuscript can be found in gzipped tar source files (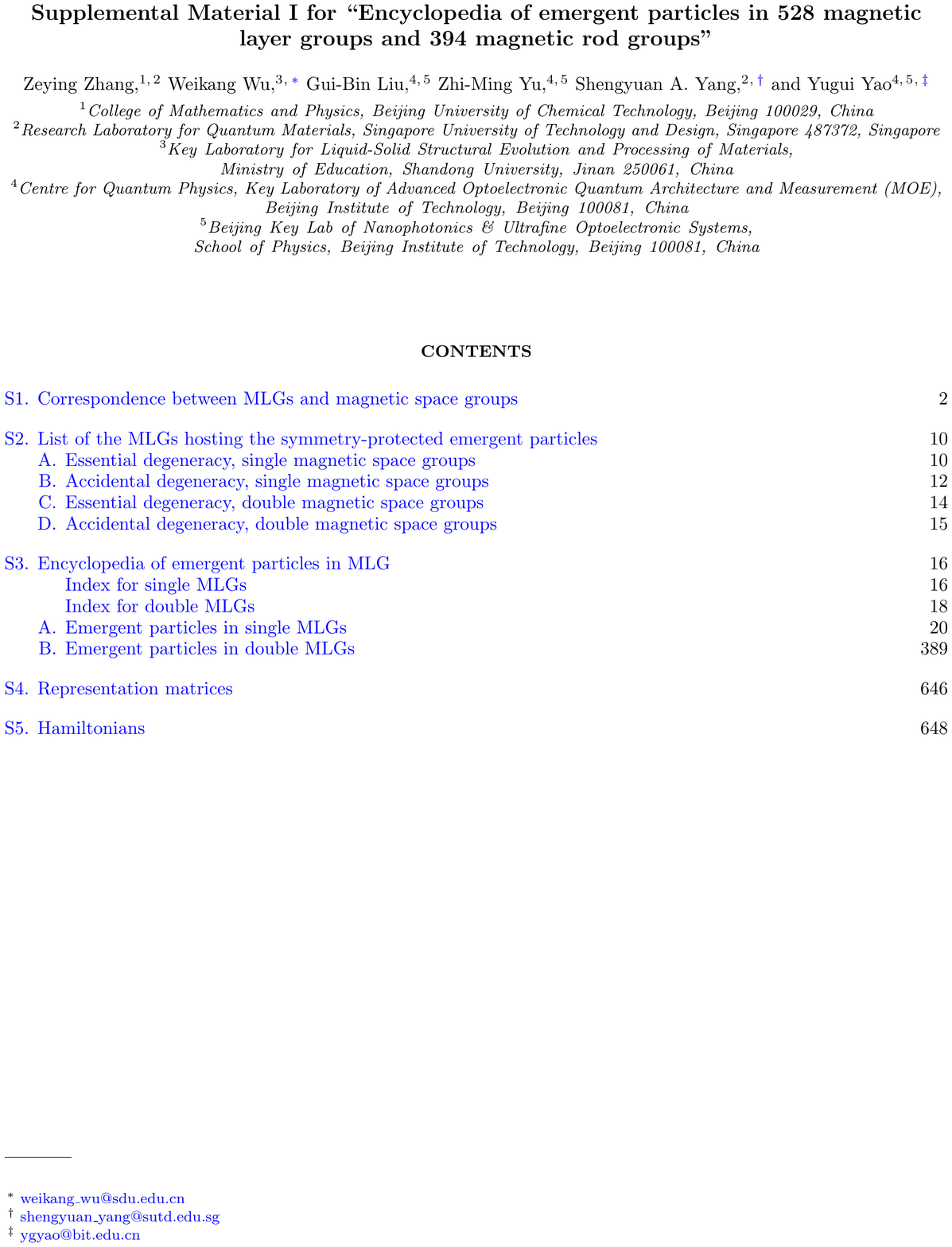 and 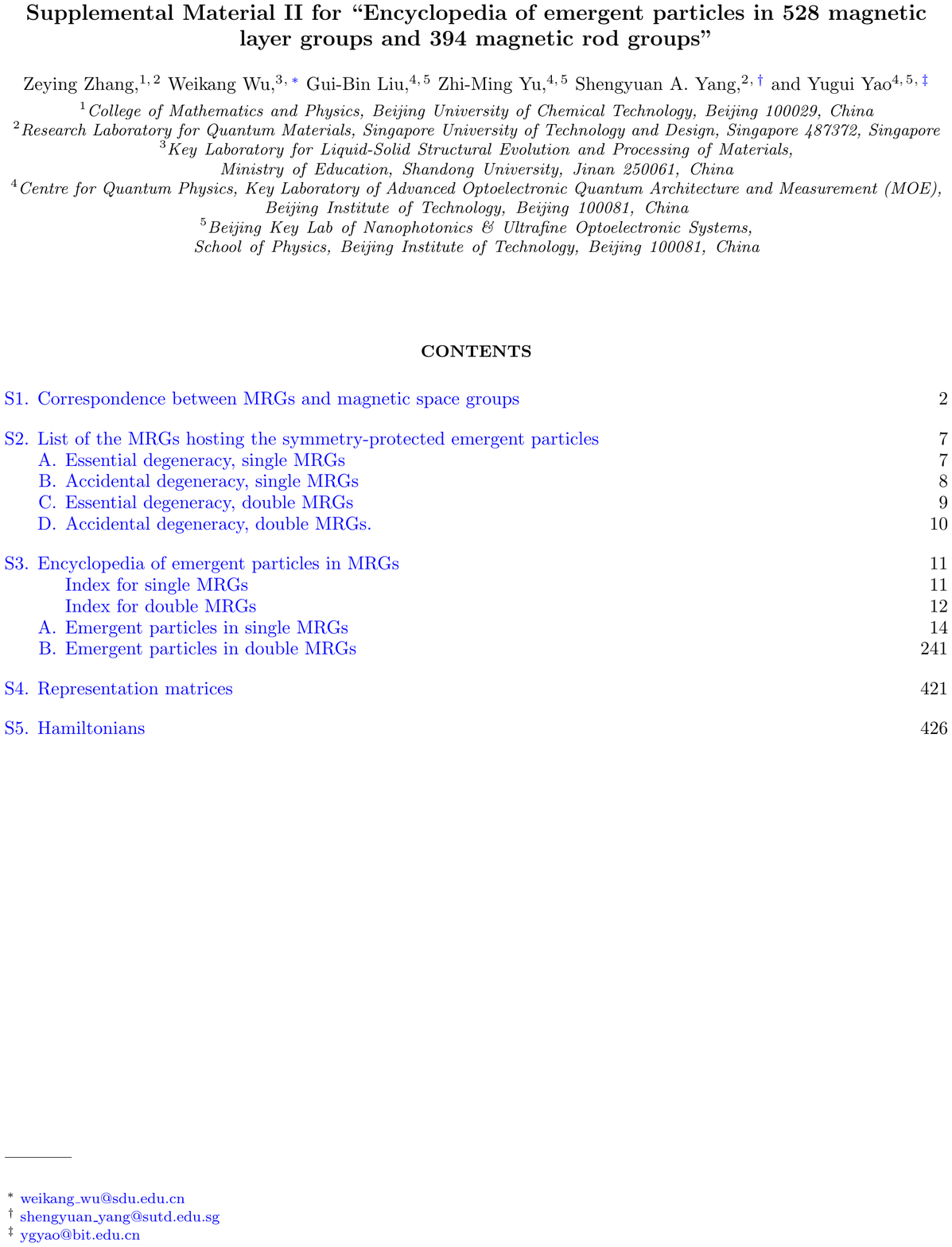).}

\bibliography{layer2}




\end{document}